\documentclass[10pt,twocolumn]{article}

\usepackage[utf8]{inputenc}
\usepackage[T1]{fontenc}
\usepackage{times}
\usepackage[margin=0.75in]{geometry}
\usepackage{booktabs}
\usepackage{tabularx}
\usepackage{array}
\usepackage{graphicx}
\usepackage{hyperref}
\usepackage{url}
\usepackage{natbib}
\usepackage{enumitem}
\usepackage{fancyvrb}
\usepackage{microtype}
\usepackage{caption}
\usepackage{float}
\usepackage{tikz}
\usetikzlibrary{arrows.meta,positioning,shapes.geometric}

\newcolumntype{L}[1]{>{\raggedright\arraybackslash}p{#1}}

\title{A Skill-Based Agentic Pipeline for Library of Congress Subject Indexing}

\author{Eric H. C. Chow\\
School of Humanities, The University of Hong Kong\\
\texttt{eric.chow@hku.hk}}

\date{}

\begin{document}
\maketitle

\begin{abstract}
This paper presents a modular AI agentic skill pipeline for automating subject indexing with Library of Congress Subject Headings (LCSH). Subject indexing---the process of analyzing a work's aboutness, selecting controlled vocabulary terms, and encoding them as MARC~21 subject access fields---is one of the most time-consuming components of library cataloging. The system decomposes this process into four discrete, sequentially executed agent skills: conceptual analysis, quantitative filtering, authority validation, and MARC field synthesis. Each skill encodes domain knowledge drawn directly from Library of Congress Subject Headings Manual (SHM) instruction sheets and subject analysis theory. The pipeline was evaluated against a corpus of ten titles whose existing subject headings were captured from the Harvard Library bibliographic dataset (a snapshot of their Alma ILS). Results demonstrate strong conceptual alignment with professional subject indexing practice, with notable differences in specificity, subdivision practice, and the agent's adherence to the 2026 LC policy discontinuing form subdivisions (\$v) in favor of LCGFT 655 fields.
\end{abstract}


\section{Introduction}

Subject indexing---the process of determining what a work is about, selecting appropriate controlled vocabulary terms, and encoding them as subject access fields in a bibliographic record---remains one of the most time-consuming and cognitively demanding components of library cataloging. It requires knowledge of a controlled vocabulary exceeding one million authorized headings, fluency with a complex set of policy documents (the Subject Headings Manual), and the professional judgment to balance specificity, depth of indexing, and end-user discoverability. The complexity of the Library of Congress Subject Headings (LCSH) system---with its elaborate subdivision rules, free-floating patterns, and policy manual spanning hundreds of instruction sheets---means that catalogers undergo years of training before they can independently construct complex subject strings, and in many cases training is not provided at all, leaving catalogers to rely on existing bibliographic records as examples \citep{Chow2024}. The resulting bottleneck has produced significant cataloging backlogs at institutions worldwide, leaving newly acquired materials undiscoverable to patrons \citep{Tang2025}.

\subsection{Literature Review}

\subsubsection{Traditional Machine Learning Approaches}

Efforts to automate subject indexing predate the current generation of large language models. National libraries have pursued machine learning solutions for over a decade, typically framing the task as an extreme multi-label text classification (XMTC) problem in which documents must be assigned labels from a vocabulary of tens or hundreds of thousands of controlled terms \citep{DSouza2026}.

The National Library of Estonia developed Kratt, a prototype automatic subject indexing tool that used page-level logistic regression classifiers to assign terms from the Estonian Subject Thesaurus \citep{Asula2021}. Kratt processed books approximately 10--15 times faster than human catalogers, completing subject indexing in about one minute per title. However, professional catalogers rated the quality of the assigned subjects as unsatisfactory, citing many inaccurate or missing terms---although a small sample of regular library users found the results somewhat more useful for discovery. The system's training data suffered from severe label sparsity: the median frequency of unique labels was only~2 across the training set, meaning most subject terms had too few examples for robust learning.

The National Library of Finland's Annif toolkit represents a more mature approach, combining multiple XMTC algorithms---including Omikuji/Bonsai (partitioned label trees), MLLM (lexical matching against controlled vocabulary terms), and XTransformer (fine-tuned BERT-style models)---into configurable ensembles \citep{Suominen2025}. Annif has been adopted by several European national libraries and supports multiple languages and vocabularies, demonstrating the adaptability required for production-scale subject classification.

The release of TIB-SID, a bilingual corpus of 136,000 catalog records annotated with GND subjects, has provided a complementary ML-ready benchmark \citep{DSouza2026}. The dataset's statistical profile reveals the fundamental challenge: subject vocabularies exhibit extreme long-tail distributions, with the vast majority of terms appearing in very few training records. This sparsity problem is structural rather than incidental---controlled vocabularies are designed to be specific, which inherently limits the number of works assignable to any given heading.

\subsubsection{Large Language Models for Subject Indexing}

The emergence of generative AI has prompted a distinct line of investigation. \citet{Brzustowicz2023} conducted early experiments testing ChatGPT's ability to generate complete MARC records. The results showed that ChatGPT could produce records comparable to those created by professional catalogers for basic metadata fields (title, author, publisher), and could even generate original records for items without existing WorldCat entries. However, notable discrepancies emerged in the assignment of subject access points, suggesting that while LLMs internalize general cataloging patterns from their training data, they lack the systematic application of LCSH policy rules required for reliable subject heading construction.

\citet{Chow2024} conducted a more focused experiment, using ChatGPT to assign LCSH to 30 electronic theses and dissertations (ETDs). The results were sobering: only approximately half of the AI-generated headings were both valid LCSH terms and sufficiently specific. The model frequently struggled with complex multi-part headings and subdivisions, producing terms that were not authorized LCSH entries---including single-word or colloquial topic descriptions. Nonetheless, the authors noted that the cost was only about \$0.25 and three minutes of processing time for all 30 documents, and that ``refining an existing (even if imperfect) AI suggestion is less daunting than constructing new subject headings from scratch'' \citep[p.~582]{Chow2024}. This framing of AI as a first-draft assistant rather than a replacement for human judgment recurs throughout the literature.

\citet{Tang2025} synthesized findings across multiple studies and reported overall poor performance for AI chatbots on LCSH assignment, with precision or F1 scores far below professional cataloging levels---one evaluation found only 26--35\% alignment between AI-generated and human-assigned LCSH. Their proposed solution is a hybrid architecture combining AI-generated candidate terms with automated validation through the Library of Congress Linked Data Service, using Model Context Protocol (MCP) integration to verify headings against authority files in real time. This approach shares conceptual ground with the present study's authority validation skill, though Tang and Jiang's system validates at a single step rather than decomposing the full cataloging workflow into multiple stages.

The most comprehensive empirical evaluation to date is the SemEval-2025 LLMs4Subjects shared task \citep{DSouza2025}, the first community benchmark specifically designed to test LLM-based subject indexing. The task challenged 14 teams to assign subjects from the German Integrated Authority File (GND)---a taxonomy of over 200,000 controlled terms---to bilingual (English/German) bibliographic records from TIB's open-access catalog. Systems were evaluated both quantitatively (precision, recall, F1 at multiple cutoffs) and qualitatively by 17 subject specialists across 28 disciplines. Participating teams deployed a range of LLM-based strategies, including retrieval-augmented generation (RAG), knowledge distillation from GND hierarchies, and multi-stage pipelines with LLM-driven re-ranking. Yet the top-performing system in the all-subjects category was Annif \citep{Suominen2025}---a traditional XMTC ensemble that used LLMs only for auxiliary preprocessing (translation and synthetic data generation), not for the core subject prediction task. A key conclusion of the shared task was that ``the advantages of LLMs over traditional machine learning algorithms for subject indexing remain debatable'' \citep[p.~2]{DSouza2025}, with smaller, well-engineered systems often rivaling large instruction-tuned LLMs. This result reinforces the pattern observed in the single-prompt studies: LLMs can identify topically relevant terms, but translating that capability into accurate, authority-controlled subject assignments remains an open challenge.

\subsubsection{Present Contribution}

The existing literature reveals a consistent pattern: AI systems can identify relevant topics in a document, but they struggle to translate those topics into properly constructed LCSH strings that comply with the Subject Headings Manual. The accuracy failures documented by \citet{Chow2024}, \citet{Brzustowicz2023}, and \citet{Tang2025} are not primarily failures of topic identification---they are failures of rule application. The models produce unauthorized heading forms, omit required subdivisions, use deprecated subfield practices, and fail to verify headings against authority files.

A key observation is that every prior LLM-based study treats subject assignment as a \emph{single-step task}. \citet{Brzustowicz2023} and \citet{Chow2024} both use direct instruction prompts that ask the model to go from a title and abstract to finished MARC fields in one pass, with no intermediate reasoning stages. None of the existing studies employ chain-of-thought (CoT) prompting or any form of structured multi-step reasoning, despite the well-documented effectiveness of CoT techniques for improving LLM performance on complex, multi-step tasks. Yet subject indexing is precisely such a task: \citet{Holley2021} describe aboutness determination and conceptual analysis as cognitive processes that \emph{precede} heading selection in professional practice. A subject cataloger does not jump from reading a title page to producing MARC fields; they move through distinct stages---identifying what a work is about, deciding which topics warrant headings, verifying those headings against authority files, and constructing properly subdivided strings. Collapsing this multi-stage cognitive process into a single computational step explains why current systems fail at rule application even when they succeed at topic identification.

This paper addresses that structural mismatch by decomposing the subject indexing workflow into discrete \emph{agent skills}---modular, reusable instruction sets that guide an LLM through one well-defined stage of the process, passing structured output to the next stage. The pipeline can be understood as an \emph{externalized, architecturally enforced} form of chain-of-thought reasoning: rather than relying on the model to internally reason through each stage, the system produces explicit intermediate outputs---a concept list, a filtered set of candidate headings, validated authority forms---that are inspectable and correctable at each step before proceeding to the next. This design provides the transparency and auditability that a single-prompt approach cannot.

Unlike the XMTC approaches of \citet{Asula2021} and \citet{Suominen2025}, the system does not require training data or corpus-level pattern learning. Unlike the single-prompt LLM approaches of \citet{Brzustowicz2023} and \citet{Chow2024}, it explicitly encodes SHM rules at each stage rather than relying on the model's internalized knowledge of LCSH conventions. And unlike the validation-only approach of \citet{Tang2025}, it decomposes the \emph{entire} subject indexing workflow---from conceptual analysis through MARC field construction---into auditable, policy-grounded stages.

The system described here was developed using Claude Code (Anthropic) and its agent skill framework. Four skills were authored by translating the normative content of eleven Library of Congress policy documents and one academic review article into machine-executable instructions. The skills were then evaluated on a corpus of ten books spanning the humanities, social sciences, sciences, and fiction.

\section{Methodology}

\subsection{Source Material and Skill Construction}

The four agent skills were constructed through close reading and systematic encoding of eleven Library of Congress policy documents, listed in Table~\ref{tab:refdocs}. These PDFs, obtained from the Library of Congress Cataloging and Acquisitions division, served as the primary normative source for the rules embedded in each skill.

\begin{table*}[t]
\centering
\caption{Library of Congress policy documents used in skill construction.}
\label{tab:refdocs}
\small
\begin{tabularx}{\textwidth}{@{}l L{8.5cm} L{4.5cm}@{}}
\toprule
\textbf{Document} & \textbf{Title / Scope} & \textbf{Primary Skill(s)} \\
\midrule
H~80  & Heading ordering by predominance & Quantitative Filtering \\
H~180 & Assigning and constructing subject headings & Conceptual Analysis, Quant.\ Filtering \\
H~405 & Establishing entities in name vs.\ subject authority & Authority Validation, MARC Synthesis \\
H~430 & Name headings as subjects & Authority Validation \\
H~830 & Geographic subdivision (indirect method) & Authority Validation, MARC Synthesis \\
H~860 & Subdivisions further subdivided by place & MARC Synthesis \\
H~1075 & Subdivisions: types, order, construction & MARC Synthesis \\
J~105 & MARC coding of LC genre/form terms & MARC Synthesis \\
J~107 & MARC authority records for genre/form terms & Authority Validation \\
J~110 & Assigning genre/form terms (LCGFT) & Quant.\ Filtering, MARC Synthesis \\
\small{LC memo} & Expand use of LCGFT and implement LCSH cataloging simplification (Jan.\ 5, 2026) & All skills \\
\bottomrule
\end{tabularx}
\end{table*}

In addition to the LC policy documents, the conceptual analysis skill drew on \citet{Holley2021}, a review article that synthesizes the theoretical frameworks for aboutness determination and conceptual analysis in library and information science. Their presentation of Wilson's four methods of subject determination (purposive, figure-ground, objective, cohesion), Langridge's three questions, and the Joudrey--Taylor concept identification framework provided the intellectual scaffolding for the first stage of the pipeline.

Each reference document was read in full and its normative content---rules, exceptions, decision criteria, and worked examples---was translated into structured instructions within a SKILL.md file. The resulting skills are not simply prompts; they are multi-section documents that define terminology, enumerate decision steps, specify output formats, and include validation checks. The complete SKILL.md files for all four skills, along with the Python scripts for authority validation, are available in the project's GitHub repository.\footnote{\url{https://github.com/choweric/subject-indexing-skills}}

\subsection{The Four-Skill Pipeline}

\begin{figure}[h]
\centering
\begin{tikzpicture}[
    node distance=0.35cm,
    inputbox/.style={
        draw, rounded corners=3pt, fill=gray!12,
        font=\sffamily\scriptsize, align=center,
        minimum height=0.7cm, minimum width=2.6cm,
        inner sep=3pt
    },
    skillbox/.style={
        draw, rounded corners=3pt, fill=blue!8,
        font=\sffamily\scriptsize, align=center,
        minimum height=1.0cm, minimum width=2.6cm,
        inner sep=3pt
    },
    arr/.style={-{Stealth[length=4pt]}, thick, gray!70}
]
\node[inputbox] (input) {Title + Abstract\\+ Table of Contents};

\node[skillbox, below=of input] (s1) {\textbf{Skill 1}\\Conceptual\\Analysis};
\node[skillbox, below=of s1] (s2) {\textbf{Skill 2}\\Quantitative\\Filtering};
\node[skillbox, below=of s2] (s3) {\textbf{Skill 3}\\Authority\\Validation};
\node[skillbox, below=of s3] (s4) {\textbf{Skill 4}\\MARC 6xx\\Synthesis};

\node[inputbox, below=of s4] (output) {MARC 6xx Fields\\(650, 600, 655, \ldots)};

\node[font=\sffamily\tiny\itshape, text=gray!70, right=0.15cm of s1.south east, anchor=north west] {concept list};
\node[font=\sffamily\tiny\itshape, text=gray!70, right=0.15cm of s2.south east, anchor=north west] {candidate headings};
\node[font=\sffamily\tiny\itshape, text=gray!70, right=0.15cm of s3.south east, anchor=north west] {validated headings};

\node[font=\sffamily\tiny, text=gray!55, left=0.15cm of s1.west, anchor=east, align=right] {Wilson, Langridge,\\Joudrey--Taylor};
\node[font=\sffamily\tiny, text=gray!55, left=0.15cm of s2.west, anchor=east, align=right] {SHM H\,80, H\,180\\20\% Rule, Rule of 3};
\node[font=\sffamily\tiny, text=gray!55, left=0.15cm of s3.west, anchor=east, align=right] {LCSH / LCGFT index\\LCNAF API};
\node[font=\sffamily\tiny, text=gray!55, left=0.15cm of s4.west, anchor=east, align=right] {H\,1075, H\,830\\J\,105, J\,110};

\draw[arr] (input) -- (s1);
\draw[arr] (s1) -- (s2);
\draw[arr] (s2) -- (s3);
\draw[arr] (s3) -- (s4);
\draw[arr] (s4) -- (output);
\end{tikzpicture}
\caption{The four-skill sequential pipeline. Each skill receives the output of the previous stage. Left annotations indicate the primary knowledge sources; right annotations show intermediate data products.}
\label{fig:pipeline}
\end{figure}
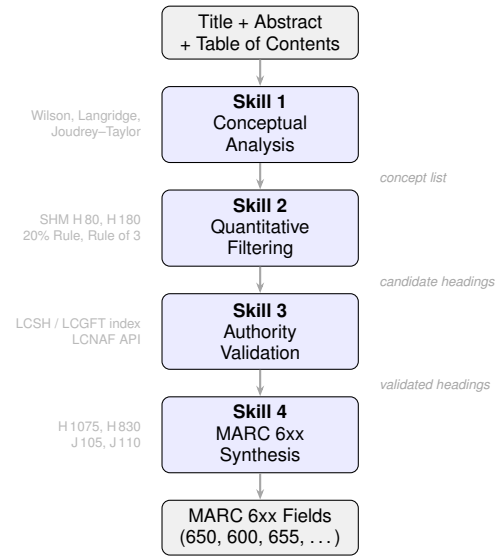

The system implements a sequential pipeline in which the output of each skill serves as input to the next (Figure~\ref{fig:pipeline}).

\textbf{Skill~1: Conceptual Analysis.} Receives the title, abstract, and table of contents of a work. Applies Wilson's four methods of subject determination (purposive, figure-ground, objective, cohesion) and Langridge's three questions (``What is it?'', ``What is it about?'', ``What is it for?'') to produce an exhaustive list of potentially significant concepts. The skill intentionally over-generates at this stage---filtering is deferred to Skill~2. The output is an aboutness statement and a flat list of candidate concepts with brief justifications.

\textbf{Skill~2: Quantitative Filtering.} Receives the concept list from Skill~1 and applies the quantitative rules codified in SHM H~180. The key decision points are:

\begin{itemize}[nosep,leftmargin=1.2em]
\item \emph{20\% Rule (H~180 \S E)}: A topic must comprise approximately 20\% or more of the work to warrant a subject heading, with exceptions for named entities critical to the work.
\item \emph{Rule of Three (H~180 \S I)}: When four or more subtopics of a broader subject are present, assign only the broader heading.
\item \emph{Specificity (H~180 \S G)}: Headings should be as specific as the content warrants.
\item \emph{Depth of indexing (H~180 \S H)}: Do not assign both a heading and a broader heading that subsumes it unless each captures a distinct facet.
\item \emph{Ordering by predominance (H~80)}: The first heading assigned should represent the primary focus of the work.
\end{itemize}

The skill also applies the 2026 LC policy change on genre/form: rather than assigning form subdivisions (\$v) within 650 fields, form concepts are routed to LCGFT 655 fields. The output is an ordered list of candidate headings with their intended MARC field type (600, 610, 650, 651, or 655).

\textbf{Skill~3: Authority Validation.} Receives the candidate headings and validates each against the appropriate LC authority file. The skill employs two lookup methods:

\begin{itemize}[nosep,leftmargin=1.2em]
\item \emph{Subjects (LCSH) and Genre/Forms (LCGFT)}: A local TF-IDF index built from NDJSON authority files downloaded from id.loc.gov. The subject index covers approximately 1.01~million authorized headings; the genre/form index covers approximately 8,600 terms.
\item \emph{Names (LCNAF)}: The LC Linked Data Service suggest2 API, queried in real time. This avoids downloading the 44~GB LCNAF authority file while providing access to over 12~million name authority records.
\end{itemize}

The skill checks whether each candidate heading exists as an authorized form, suggests the authorized form when a variant is detected (via UF references), and verifies geographic subdivision authorization per H~830.

\textbf{Skill~4: MARC 6xx Synthesis.} Receives the validated headings and constructs properly formatted MARC~21 subject access fields. The skill implements the subdivision order prescribed in H~1075 (\$a~$\rightarrow$~\$z~$\rightarrow$~\$x~$\rightarrow$~\$y), applies the indirect geographic subdivision method (H~830), and enforces the 2026 discontinuation of \$v form subdivisions. For name headings (600, 610, 611), the skill applies H~405 rules for authority file determination and H~430 validation criteria.

\subsection{Test Corpus}

Ten titles were selected to represent a range of disciplines, publication types, and subject indexing challenges (Table~\ref{tab:corpus}). The titles were sampled from the Harvard Library Bibliographic Dataset \citep{HarvardLibrary2022}, an open-access collection of over 12.7~million MARC~XML bibliographic records exported from Harvard Library's Alma integrated library system. The dataset, released pursuant to Harvard Library's Open Metadata Policy, provides a snapshot of all active bibliographic records with one or more active holdings. For each sampled title, we extracted the existing LCSH assignments from the Harvard record's 6XX fields; these human-assigned headings serve as the comparison baseline, yielding ten paired comparisons.

Bibliographic metadata for the ten titles---including title, abstract, table of contents, ISBN, and publisher information---was assembled in a spreadsheet and is available as a supplementary Excel file in the project's GitHub repository alongside the SKILL.md files and pipeline output transcripts.\footnote{\url{https://github.com/choweric/subject-indexing-skills}}

\begin{table*}[t]
\centering
\caption{Test corpus sampled from the Harvard Library Bibliographic Dataset.}
\label{tab:corpus}
\small
\begin{tabularx}{\textwidth}{@{}c L{9.2cm} L{3.2cm} L{2.6cm}@{}}
\toprule
\textbf{\#} & \textbf{Title} & \textbf{Author / Editor} & \textbf{Discipline} \\
\midrule
1  & Classical and Christian ideas in English Renaissance poetry: a student's guide & Rivers, I.\ (1994) & Literary criticism \\
2  & Trauma room two & Green, P.~A.\ (2015) & Emergency med.\ / Fiction \\
3  & The color line and the assembly line: managing race in the Ford empire & Esch, E.~D.\ (2018) & Labor history / Race studies \\
4  & Banker to the poor: micro-lending and the battle against world poverty & Yunus, M.\ (2003) & Economics / Development \\
5  & Currencies, commodities and consumption & Clements, K.~W.\ (2013) & Economics \\
6  & Respiratory virology and immunogenicity & Pokorski, M.\ (Ed.) (2015) & Biomedical sciences \\
7  & Artificial black holes & Volovik, G.\ (Ed.) (2002) & Physics \\
8  & The hallowing of logic: the Trinitarian method of Richard Baxter's \emph{Methodus theologiae} & Burton, S.~J.~G.\ (2012) & Theology / Hist.\ of ideas \\
9  & All joking aside: American humor and its discontents & Krefting, R.\ (2014) & Cultural studies / Humor \\
10 & Deference revisited: Andean ritual in the plurinational state & Goudsmit, I.~A.\ (2016) & Anthropology \\
\bottomrule
\end{tabularx}
\end{table*}

\subsection{Evaluation Method}

For each of the ten paired titles, the agent pipeline's MARC output was compared with the Harvard/LC headings along four dimensions: (1)~conceptual recall---did the agent identify the same broad topics as the human indexer? (2)~heading precision---did the agent select the correct authorized heading form? (3)~subdivision accuracy---did the agent construct subdivisions (\$x, \$z, \$y) correctly? (4)~genre/form treatment---did the agent assign LCGFT 655 fields appropriately?

Because the baseline headings were captured from a live ILS that may reflect subject indexing decisions made years or decades ago---potentially under different LC policies---differences are not automatically errors. The comparison is interpretive rather than strictly quantitative.

\section{Results}

\subsection{Heading-by-Heading Comparison}

Table~\ref{tab:comparison} presents the complete heading-by-heading comparison for all ten paired titles. Agent-generated headings are shown in standard MARC notation; Harvard/LC headings are shown as they appeared in the bibliographic dataset, with subdivisions delimited by ``\,--\,''. Due to space constraints, abbreviated analyses are provided for each title; full analyses are available in the extended version of this paper.

\begin{table*}[p]
\centering
\caption{Agent output vs.\ Harvard/LC headings for all ten titles.}
\label{tab:comparison}
\scriptsize
\setlength{\tabcolsep}{3pt}
\renewcommand{\arraystretch}{0.95}
\begin{tabularx}{\textwidth}{@{}L{0.58\textwidth} L{0.38\textwidth}@{}}
\toprule
\textbf{Agent Output} & \textbf{Harvard/LC Headings} \\
\midrule
\multicolumn{2}{@{}l}{\textbf{Title 1: \emph{Classical and Christian Ideas in English Renaissance Poetry} (Rivers, 1994)}} \\
\midrule
\texttt{650\,\#0 \$aEnglish poetry\$yEarly modern, 1500-1700\$xHistory and criticism.} & English poetry -- Early modern, 1500-1700 -- History and criticism \\
\texttt{650\,\#0 \$aChristianity and literature\$zEngland.} & Christianity and literature \\
\texttt{650\,\#0 \$aClassical literature\$xAppreciation\$zEngland.} & English poetry -- Classical influences \\
\texttt{655\,\#7 \$aPrimary sources.\$2lcgft} & Christian poetry, English -- Hist.\ and criticism \\
--- & Renaissance -- England \\
--- & Classicism -- England \\
\midrule
\multicolumn{2}{@{}l}{\textbf{Title 2: \emph{Trauma Room Two} (Green, 2015)}} \\
\midrule
\texttt{650\;\;0 \$aEmergency medicine.} & Trauma centers -- Fiction \\
\texttt{650\;\;0 \$aHospitals\$xEmergency services.} & Emergency physicians -- Fiction \\
\texttt{650\;\;0 \$aEmergency physicians.} & Physicians -- Fiction \\
\texttt{650\;\;0 \$aPhysician and patient.} & Short stories, American \\
\texttt{655\;\;7 \$aShort stories.\$2lcgft} & Autobiographical fiction, American \\
\texttt{655\;\;7 \$aMedical fiction.\$2lcgft} & --- \\
\midrule
\multicolumn{2}{@{}l}{\textbf{Title 3: \emph{The Color Line and the Assembly Line} (Esch, 2018)}} \\
\midrule
\texttt{610\,20 \$aFord Motor Company\$xEmployees\$xHistory\$y20th century.} & Automobile industry workers -- Social conditions -- 20th century \\
\texttt{650\;\;0 \$aRace discrimination\$zUnited States\$xHistory\$y20th century.} & Racism in the workplace -- Michigan -- 20th century \\
\texttt{650\;\;0 \$aRace discrimination\$zBrazil\$xHistory\$y20th century.} & Racism in the workplace -- Brazil -- 20th century \\
\texttt{650\;\;0 \$aRace discrimination\$zSouth Africa\$xHistory\$y20th century.} & Racism in the workplace -- South Africa -- Port Elizabeth -- 20th century \\
\texttt{650\;\;0 \$aAfrican Americans\$xEmployment\$xHistory\$y20th century.} & --- \\
\texttt{650\;\;0 \$aFordism\$xHistory.} & --- \\
\texttt{655\;\;7 \$aCase studies.\$2lcgft} & --- \\
\midrule
\multicolumn{2}{@{}l}{\textbf{Title 4: \emph{Banker to the Poor} (Yunus, 2003)}} \\
\midrule
\texttt{600\,10 \$aYunus, Muhammad,\$d1940-} & Economists -- Bangladesh -- Biography \\
\texttt{610\,20 \$aGrameen Bank.} & Banks and banking -- Bangladesh \\
\texttt{650\;\;0 \$aMicrofinance.} & Microfinance -- Bangladesh -- History \\
\texttt{650\;\;0 \$aPoverty.} & Rural poor -- Bangladesh -- History \\
\texttt{655\;\;7 \$aAutobiographies.\$2lcgft} & --- \\
\midrule
\multicolumn{2}{@{}l}{\textbf{Title 5: \emph{Currencies, Commodities and Consumption} (Clements, 2013)}} \\
\midrule
\texttt{650\;\;0 \$aPurchasing power parity.} & Purchasing power parity \\
\texttt{650\;\;0 \$aForeign exchange rates.} & Foreign exchange \\
\texttt{650\;\;0 \$aPrices.} & Consumer price indexes \\
\texttt{650\;\;0 \$aConsumption (Economics).} & Cost and standard of living \\
\midrule
\multicolumn{2}{@{}l}{\textbf{Title 6: \emph{Respiratory Virology and Immunogenicity} (Pokorski, 2015)}} \\
\midrule
\texttt{650\;\;0 \$aInfluenza vaccines.} & Medicine \\
\texttt{650\;\;0 \$aInfluenza.} & Immunology \\
\texttt{650\;\;0 \$aImmune response.} & Vaccines \\
\texttt{650\;\;0 \$aRespiratory infections.} & Medical virology \\
\texttt{650\;\;0 \$aInfluenza vaccines\$zPoland.} & --- \\
\texttt{655\;\;7 \$aEssays.\$2lcgft} & --- \\
\midrule
\multicolumn{2}{@{}l}{\textbf{Title 7: \emph{Artificial Black Holes} (Volovik, 2002)}} \\
\midrule
\texttt{650\,\#0 \$aBlack holes (Astronomy)} & Black holes (Astronomy) -- Mathematical models \\
\texttt{650\,\#0 \$aGeneral relativity (Physics)} & Quantum gravity -- Mathematical models \\
\texttt{650\,\#0 \$aCondensed matter.} & Condensed matter physics -- Mathematics \\
\texttt{650\,\#0 \$aSuperfluidity.} & --- \\
\texttt{655\,\#7 \$aEssays.\$2lcgft} & --- \\
\midrule
\multicolumn{2}{@{}l}{\textbf{Title 8: \emph{The Hallowing of Logic} (Burton, 2012)}} \\
\midrule
\texttt{600\,10 \$aBaxter, Richard,\$d1615-1691.\$tMethodus theologiæ Christianæ.} & Theology -- Methodology \\
\texttt{650\;\;0 \$aTrinity.} & Trinity -- History of doctrines \\
\texttt{650\;\;0 \$aTheology\$xMethodology.} & Salvation -- Christianity -- History of doctrines \\
\texttt{650\;\;0 \$aProtestant Scholasticism.} & --- \\
\midrule
\multicolumn{2}{@{}l}{\textbf{Title 9: \emph{All Joking Aside} (Krefting, 2014)}} \\
\midrule
\texttt{650\;\;0 \$aStand-up comedy\$zUnited States\$xPolitical aspects.} & Stand-up comedy -- United States \\
\texttt{650\;\;0 \$aStand-up comedy\$zUnited States\$xSocial aspects.} & Comedy -- History and criticism \\
\texttt{650\;\;0 \$aAmerican wit and humor\$xHistory and criticism.} & Participatory theater \\
\texttt{650\;\;0 \$aWomen comedians\$zUnited States.} & --- \\
\texttt{600\,10 \$aTyler, Robin.} & --- \\
\texttt{600\,10 \$aKondabolu, Hari.} & --- \\
\midrule
\multicolumn{2}{@{}l}{\textbf{Title 10: \emph{Deference Revisited: Andean Ritual in the Plurinational State} (Goudsmit, 2016)}} \\
\midrule
\texttt{650\,\#0 \$aIndians of South America\$zBolivia\$xRites and ceremonies.} & Indians of South America -- Bolivia -- Toracari -- Government relations \\
\texttt{650\,\#0 \$aIndians of South America\$zBolivia\$xPolitics and government.} & Indians of South America -- Bolivia -- Toracari -- Social life and customs \\
\texttt{650\,\#0 \$aLandlord and tenant\$zBolivia.} & Peasants -- Bolivia -- Toracari -- Government relations \\
\texttt{650\,\#0 \$aPolitical anthropology\$zBolivia.} & Peasants -- Bolivia -- Toracari -- Social life and customs \\
\texttt{651\,\#0 \$aBolivia\$xPolitics and government\$y2006-} & Landlords -- Bolivia -- Toracari \\
\texttt{655\,\#7 \$aEthnographies.\$2lcgft} & Land tenure -- Bolivia -- Toracari \\
\bottomrule
\end{tabularx}
\end{table*}

\textbf{Title~1} (\emph{Renaissance Poetry}). The agent's first heading is an exact match. Its second heading adds a geographic subdivision (\$zEngland) not present in the baseline record but defensible given the work's exclusive focus on England. The agent assigned fewer headings overall (3~vs.~6), omitting ``Christian poetry, English,'' ``Renaissance,'' and ``Classicism.'' The LCGFT assignment (\emph{Primary sources}) reflects the work's inclusion of source text extracts.

\textbf{Title~2} (\emph{Trauma Room Two}). This title exposes a fundamental difference in fiction cataloging. The baseline record uses topical headings with the form subdivision ``Fiction'' appended (pre-2026 practice), while the agent correctly applied the 2026 policy: topical headings without form subdivisions, plus separate LCGFT 655 fields (\emph{Short stories}, \emph{Medical fiction}).

\textbf{Title~3} (\emph{Color Line}). The baseline record used the more specific \emph{Racism in the workplace} while the agent used the broader \emph{Race discrimination}. The baseline headings localized to Michigan and Port~Elizabeth; the agent used country-level geography throughout. The agent added a corporate name heading (\emph{Ford Motor Company}) and the concept \emph{Fordism}, neither of which appears in the baseline.

\textbf{Title~4} (\emph{Banker to the Poor}). The agent expressed the biographical dimension through a 600 name heading (\emph{Yunus, Muhammad}) and a 655 LCGFT field (\emph{Autobiographies}), while the baseline record used a form subdivision (\emph{Biography}). The baseline headings include geographic and chronological subdivisions; the agent's headings are less subdivided. The agent used \emph{Poverty} where the baseline used the more specific \emph{Rural poor}.

\textbf{Title~5} (\emph{Currencies, Commodities}). The closest match in the corpus. \emph{Purchasing power parity} is an exact match. The remaining three headings show minor synonym-level variation typical of inter-cataloger disagreement.

\textbf{Title~6} (\emph{Respiratory Virology}). The agent and baseline diverge significantly in specificity. The baseline record used very broad headings (\emph{Medicine}, \emph{Immunology}); the agent's headings are markedly more specific (\emph{Influenza vaccines}, \emph{Immune response}). Per SHM H~180's specificity principle, the agent's approach is arguably more consistent with LC policy.

\textbf{Title~7} (\emph{Artificial Black Holes}). The baseline record added the subdivision \emph{Mathematical models} to reflect the analog-modeling focus; the agent omitted this dimension. The baseline headings better capture the \emph{modeling} dimension, while the agent provided broader topical coverage including \emph{Superfluidity}.

\textbf{Title~8} (\emph{Hallowing of Logic}). The agent produced a sophisticated 600 name-title construction (\emph{Baxter, Richard\ldots\$tMethodus theologiæ Christianæ}), absent from the baseline record. Both include \emph{Theology -- Methodology} and \emph{Trinity}. The agent added \emph{Protestant Scholasticism}; the baseline included \emph{Salvation -- Christianity -- History of doctrines}.

\textbf{Title~9} (\emph{All Joking Aside}). Both the agent and baseline identified stand-up comedy as the primary subject. The agent subdivided into \emph{Political aspects} and \emph{Social aspects} and assigned 600 fields for individual comedians (Tyler, Kondabolu). The baseline's \emph{Participatory theater} captures a facet the agent missed.

\textbf{Title~10} (\emph{Deference Revisited}). The richest comparison. The agent now uses the subdivision \emph{Politics and government} (matching the baseline's base concept) and adds a 651 geographic heading with a chronological subdivision (\emph{Bolivia\$xPolitics and government\$y2006-}), correctly situating the work in the Morales-era plurinational state period---a level of chronological specificity absent from the baseline record. The agent also assigned \emph{Political anthropology\$zBolivia}, capturing the disciplinary framework. However, the Toracari gap persists: the baseline consistently subdivides to the field-site level while the agent stops at the country level. The baseline doubles headings across two subject groups (\emph{Indians of South America} and \emph{Peasants}); the agent uses only one. The baseline includes \emph{Land tenure}, which the agent dropped in favor of the disciplinary heading. The agent chose \emph{Rites and ceremonies} (more specific) where the baseline used \emph{Social life and customs} (broader).

\subsection{Patterns Observed}

\textbf{Specificity.} In seven of ten cases, the agent produced headings at a comparable or more specific level than the baseline record. The most striking example is the medical virology title (\#6), where the agent assigned influenza-specific headings while the baseline used broad disciplinary terms. In two cases (Ford, black holes), the baseline was clearly more specific---particularly in geographic granularity and topical subdivisions. The Bolivia title presents a mixed picture: the agent added a chronological subdivision (\emph{\$y2006-}) and a disciplinary heading (\emph{Political anthropology}) absent from the baseline, but the baseline was more specific geographically (Toracari valley vs.\ country level) and topically (\emph{Land tenure}, which the agent dropped).

\textbf{Name Headings.} The agent assigned name headings (600, 610) in four of ten titles---for biographical subjects, corporate bodies, and individuals treated at length in case-study chapters. The baseline records contained no 600 or 610 fields for any title, likely reflecting differences in cataloging depth or institutional practice.

\textbf{LCGFT and the 2026 Policy.} The agent consistently applied the February 2026 LC policy change, assigning genre/form concepts in 655 fields with LCGFT vocabulary rather than as \$v form subdivisions. The baseline records use the older practice (e.g., \emph{Fiction} appended to topical headings). This is the single largest systematic difference and reflects genuine policy evolution rather than agent error.

\textbf{Subdivision Practice.} The baseline records tend to include more subdivisions per heading, particularly geographic. The Bolivia title illustrates this most clearly: the baseline drills down to the Toracari valley as a third-level geographic subdivision, while the agent stops at the country level. Notably, the agent did produce a chronological subdivision in its 651 heading (\emph{Bolivia\$xPolitics and government\$y2006-}) that the baseline record lacks, demonstrating that the pipeline can apply period subdivisions when the temporal scope is explicit. The persistent geographic gap suggests the synthesis skill may benefit from stronger guidance on when to subdivide below the country level---particularly for works with a clearly identified field site or locality.

\textbf{Conceptual Framing.} In several cases, the agent and baseline record identified the same concept through different authorized headings: \emph{Race discrimination} vs.\ \emph{Racism in the workplace}, \emph{Poverty} vs.\ \emph{Rural poor}, \emph{Landlord and tenant} vs.\ \emph{Landlords}. These differences fall within normal inter-cataloger variation.

\textbf{Subject Group Coverage.} The Bolivia title revealed that the baseline record assigned headings under two parallel subject groups (\emph{Indians of South America} and \emph{Peasants}), capturing both ethnic and socioeconomic dimensions. The agent used only one group, suggesting the conceptual analysis skill could benefit from guidance on identifying overlapping population groups.

\section{Discussion}

\subsection{Strengths of the Agent Skill Approach}

The modular pipeline design offers several advantages over monolithic prompting:

\textbf{Separation of concerns.} Each skill addresses one phase of the subject indexing process, mirroring the cognitive stages a trained subject cataloger moves through. This decomposition makes each stage auditable: a reviewer can identify exactly where a questionable heading decision originated.

\textbf{Policy updateability.} When LC policy changes---as it did in February 2026 with the discontinuation of form subdivisions---the change can be implemented by modifying the relevant skill(s) without rebuilding the entire system. The 2026 LCGFT transition required updates to two skills while leaving the other two unchanged.

\textbf{Transparency of reasoning.} Each skill produces intermediate output with justifications (20\% calculations, authority match scores, subdivision authorization checks), making the system's reasoning inspectable---in contrast to end-to-end systems that produce headings without explanation.

\textbf{Standards fidelity.} The skill instructions directly encode SHM rules with section-level granularity. The 20\% rule, Rule of Three, specificity principle, and subdivision ordering rules are not learned from training data but explicitly specified, ensuring that every heading decision is traceable to a specific policy document.

\subsection{Limitations and Areas for Improvement}

\textbf{Geographic and chronological subdivision.} The agent under-subdivided in several cases. Strengthening the MARC synthesis skill's guidance on when to append \$z and \$y subdivisions---particularly for works with clear geographic or temporal scope---would address this gap.

\textbf{Heading selection among synonymous terms.} When multiple authorized headings cover overlapping territory (e.g., \emph{Race discrimination} vs.\ \emph{Racism in the workplace}), the agent sometimes selected the broader term. Enhanced BT/NT navigation in the authority validation skill could guide toward the most specific applicable heading.

\textbf{Absence of copy cataloging context.} In practice, subject catalogers consult existing records in shared utilities like WorldCat when assigning headings. The agent pipeline operates from the title page and abstract alone, without access to peer records. Integrating a step that queries existing bibliographic records for the same ISBN could improve consistency with community subject indexing practice.

\textbf{Corpus size and disciplinary coverage.} Ten paired comparisons are insufficient for statistical generalization. Although the test corpus spans the humanities, social sciences, natural sciences, and fiction, with only one or two titles per discipline it is not possible to determine whether the pipeline performs more reliably in some subject domains than others. A more systematic evaluation---drawing a larger, stratified sample across disciplines from the Harvard Library Bibliographic Dataset---would allow researchers to identify whether the pipeline is better suited to certain types of material (e.g., monographs with well-defined topical scope vs.\ interdisciplinary edited volumes) and whether particular disciplines present recurring challenges for automated subject indexing, such as the granular geographic subdivisions observed in the anthropology title or the ``Mathematical models'' subdivisions expected in physics. Establishing inter-cataloger agreement baselines for the same titles would further contextualize the pipeline's performance relative to the inherent variability of human subject indexing.

\subsection{Comparison Baseline Considerations}

The Harvard Library bibliographic dataset serves as a convenient but imperfect baseline. Subject headings in a production ILS may reflect varying levels of cataloging (full, core, minimal), copy cataloging from diverse sources with differing subject indexing practices, legacy headings predating current SHM instructions, and batch processing or automated enrichment. The baseline headings should therefore be understood as \emph{one professional interpretation} rather than a gold standard.

\section{Conclusion}

This study demonstrates that a modular agent skill pipeline can produce LCSH subject heading assignments that are conceptually aligned with professional subject indexing practice. The system's strengths---specificity, named entity identification, and adherence to current LC policy---complement its weaknesses in subdivision completeness and synonym navigation. The agent skill architecture offers a practical framework for encoding complex subject indexing rules in a maintainable, auditable, and updateable form. It is important to note that the pipeline addresses only the subject indexing component of cataloging---the assignment of MARC 6XX subject access fields---and does not attempt descriptive cataloging, classification, or authority control for names and titles. Subject indexing is, however, widely recognized as one of the most time-consuming and expertise-intensive parts of the cataloging workflow, making it a high-value target for AI-assisted automation. Two practical use cases are particularly compelling. First, the pipeline could be used to \emph{enhance existing bibliographic records} that lack subject headings or carry only minimal-level headings---a common situation for vendor-supplied records and older catalog entries that predate current SHM practice. Second, it could help libraries \emph{efficiently process cataloging backlogs}: collections of newly acquired or donated materials that remain undiscoverable to patrons because subject indexing has not yet been performed. In both cases, the pipeline's output would serve as a draft for review by a subject cataloger rather than as a final product, consistent with the human-in-the-loop workflow advocated throughout the literature \citep{Chow2024,Tang2025}.

The most significant finding may be structural rather than evaluative: the decomposition of subject indexing into discrete, policy-grounded skills appears to be a viable paradigm for applying LLMs to complex knowledge work. Each skill is small enough to be validated against its source policy document, yet the pipeline as a whole produces output that engages with the full complexity of LCSH assignment. Where prior work has shown that single-prompt LLM approaches achieve only 26--35\% alignment with human-assigned LCSH \citep{Tang2025} and struggle with subdivision construction \citep{Chow2024}, the modular skill approach produced output with conceptual overlap on over half of all heading comparisons---suggesting that explicit rule encoding substantially improves upon reliance on internalized model knowledge alone. Future work should expand the evaluation corpus, integrate copy cataloging consultation, and explore the system's applicability to other controlled vocabularies (MeSH, AAT, FAST).



\begin{thebibliography}{99}

\bibitem[Asula et~al., 2021]{Asula2021}
Asula, M., Makke, J., Freienthal, L., Kuulmets, H.-A., \& Sirel, R. (2021).
\newblock Kratt: Developing an automatic subject indexing tool for the National Library of Estonia.
\newblock \textit{Cataloging \& Classification Quarterly}, \textit{59}(8), 775--793.
\newblock \url{https://doi.org/10.1080/01639374.2021.1998283}

\bibitem[Brzustowicz, 2023]{Brzustowicz2023}
Brzustowicz, R. (2023).
\newblock From ChatGPT to CatGPT: The implications of artificial intelligence on library cataloging.
\newblock \textit{Information Technology and Libraries}, \textit{42}(3).
\newblock \url{https://doi.org/10.5860/ital.v42i3.16295}

\bibitem[Chow et~al., 2024]{Chow2024}
Chow, E. H. C., Kao, T. J., \& Li, X. (2024).
\newblock An experiment with the use of ChatGPT for LCSH subject assignment on electronic theses and dissertations.
\newblock \textit{Cataloging \& Classification Quarterly}, \textit{62}(5), 574--588.
\newblock \url{https://doi.org/10.1080/01639374.2024.2394516}

\bibitem[Harvard Library, 2022]{HarvardLibrary2022}
Harvard Library. (2022).
\newblock \textit{Harvard Library bibliographic metadata} [Data set].
\newblock Harvard Dataverse.
\newblock \url{https://doi.org/10.7910/DVN/I8L0ZZ}

\bibitem[Holley \& Joudrey, 2021]{Holley2021}
Holley, R. M., \& Joudrey, D. N. (2021).
\newblock Aboutness and conceptual analysis: A review.
\newblock \textit{Cataloging \& Classification Quarterly}, \textit{59}(2--3), 159--185.
\newblock \url{https://doi.org/10.1080/01639374.2020.1856992}

\bibitem[D'Souza et~al., 2025]{DSouza2025}
D'Souza, J., Sadruddin, S., Israel, H., Begoin, M., \& Slawig, D. (2025).
\newblock SemEval-2025 Task 5: LLMs4Subjects---LLM-based automated subject tagging for a national technical library's open-access catalog.
\newblock \textit{arXiv preprint arXiv:2504.07199}.
\newblock \url{https://arxiv.org/abs/2504.07199}

\bibitem[D'Souza et~al., 2026]{DSouza2026}
D'Souza, J., Sadruddin, S., K\"{a}hler, M., Salfinger, A., Zaccagna, L., Incitti, F., Snidaro, L., \& Suominen, O. (2026).
\newblock An extreme multi-label text classification (XMTC) library dataset: What if we took ``Use of Practical AI in Digital Libraries'' seriously?
\newblock \textit{arXiv preprint arXiv:2603.10876}.
\newblock \url{https://arxiv.org/abs/2603.10876}

\bibitem[Suominen et~al., 2025]{Suominen2025}
Suominen, O., Inkinen, J., \& Lehtinen, M. (2025).
\newblock Annif at SemEval-2025 Task 5: Traditional XMTC augmented by LLMs.
\newblock \textit{arXiv preprint arXiv:2504.19675}.
\newblock \url{https://arxiv.org/abs/2504.19675}

\bibitem[Tang \& Jiang, 2025]{Tang2025}
Tang, K.-L., \& Jiang, Y. (2025).
\newblock Better recommendations: Validating AI-generated subject terms through LOC Linked Data Service.
\newblock \textit{arXiv preprint arXiv:2508.00867}.
\newblock \url{https://arxiv.org/abs/2508.00867}

\end{thebibliography}
\end{document}